\documentclass{Interspeech}
\usepackage{hyperref}
\usepackage{tikz}
\usepackage{amsmath}
\usepackage{graphicx}
\usepackage{textgreek}
\usepackage{algorithm}
\usepackage{algorithmic}
\usetikzlibrary[patterns]
\newcommand{\mast}{MAST}
\newcommand{\mact}{MAcT}
\newcommand{\smact}{SMAcT}

\interspeechcameraready

\title{Robust fine-tuning of speech recognition models via model merging: application to disordered speech}

\author[affiliation={1}]{Alexandre}{Ducorroy}
\author[affiliation={1}]{Rachid}{Riad}

\affiliation{}{Callyope}{France}
\email{rachid@callyope.com}

\usepackage{comment}

\begin{document}

\maketitle

\begin{abstract}
Automatic Speech Recognition (ASR) has advanced with Speech Foundation Models (SFMs), yet performance degrades on dysarthric speech due to variability and limited data. This study as part of the submission to the Speech Accessibility challenge, explored model merging to improve ASR generalization using Whisper as the base SFM. We compared fine-tuning with single-trajectory merging, combining models from one fine-tuning path, and multi-run merging, merging independently trained models. 
Our best multi-run merging approach achieved a 12\% relative decrease of WER over classic fine-tuning, and a 16.2\% relative decrease on long-form audios, a major loss contributor in dysarthric ASR.
Merging more and more models led to continuous gains, remained effective in low-data regimes, and generalized across model architectures. These results highlight model merging as an easily replicable adaptation method that consistently improves ASR without additional inference cost or hyperparameter tuning.

\end{abstract}

\section{Introduction}

Automatic Speech Recognition (ASR) has evolved from a specialized tool confined to speech labs and research environments into a mainstream technology that can be deployed with just a few lines of code\footnote{\url{https://docs.nvidia.com/nemo-framework/user-guide/24.07/nemotoolkit/asr/intro.html\#transcribe-speech-with-3-lines-of-code}}. This transformation has been driven by the development of Speech Foundation Models (SFMs), which are trained on large and diverse datasets \cite{fang2022data, hsu2021hubert, chen2022wavlm, baevski2020wav2vec}. These models achieve state-of-the-art performance on major benchmarks and demonstrate remarkable adaptability to new domains \cite{fan2022towards, fan2024benchmarking}. Furthermore, they can be fine-tuned and repurposed for various audio \cite{gong23d_interspeech}, paralinguistic \cite{goron2024improving} tasks, even healthcare applications \cite{cui2023transferring, riad2024automated} extending their utility beyond traditional ASR applications.
\begin{figure}[!ht]
    \centering
\begin{tikzpicture}
        \fill[gray!10] (-7.5,3.5) rectangle (0,0.5);
        \node[right] at (-7.5,4.6) [align=center] {\footnotesize{Pretrained model}};
        \node[right] at (-7.5, 2) [align=center] {\footnotesize{Fine tuned models }};
        \node[right] at (-7.5,0) [align=center] {\footnotesize{Final model}};

        \node at (-2.9,4.6) [align=center] {Whisper \\ (WER=33.7)};
        \node (pretrain) at (-2.9,4) [draw, minimum size=0.1cm] {}; 
        \node (vanilla1) at (-5,3) [draw, minimum size=0.1cm] {};
        \node (vanilla2) at (-5, 0) [draw, minimum size=0.1cm] {};
        \draw[->, thick] (pretrain) -- (vanilla1);
        \draw[->, thick] (vanilla1) -- (vanilla2);
        
        \node (wise1) at (-3.5,3) [draw, minimum size=0.1cm] {};
        \node (wise2) at (-3.5,2) [draw, minimum size=0.1cm] {};
        \node (wise3) at (-3.5,1.5) [draw, minimum size=0.1cm] {};
        \node (wise4) at (-3.5,1) [draw, minimum size=0.1cm] {};
        \node (wise5) at (-3.5,0) [draw, minimum size=0.1cm] {};
        \draw[->, thick] (pretrain) -- (wise1);
        \draw[->, thick] (wise1) -- (wise2);
        \draw[->, thick] (wise2) -- (wise3);
        \draw[->, thick] (wise3) -- (wise4);
        \draw[->, dashed] (wise4) -- (wise5);
        \draw[->, dashed] (wise3) to[out=-10, in=-5] (wise5);
        \draw[->, dashed] (wise2) to[out=-10, in=-5] (wise5);
        \draw[->, dashed] (wise1) to[out=-10, in=-5] (wise5);

        \node (mtraj1-0) at (-2,3) [draw, minimum size=0.1cm] {};
        \node (mtraj2-0) at (-1.4,3) [draw, minimum size=0.1cm] {};
        \node (mtraj3-0) at (-0.8,3) [draw, minimum size=0.1cm] {};
        \node (mtraj_avg) at (-1.4,0) [draw, minimum size=0.1cm] {};

        \node (mtraj1-1) at (-2,2) [draw, minimum size=0.1cm] {};
        \node (mtraj1-2) at (-2,1.5) [draw, pattern=north east lines, minimum size=0.1cm] {};
        \node (mtraj1-3) at (-2,1) [draw, minimum size=0.1cm] {}; 

        \node (mtraj2-1) at (-1.4,2) [draw, minimum size=0.1cm] {};
        \node (mtraj2-2) at (-1.4,1.5) [draw, minimum size=0.1cm] {};
        \node (mtraj2-3) at (-1.4,1) [draw, pattern=north east lines, minimum size=0.1cm] {};
        
        \node (mtraj3-1) at (-0.8,2) [draw, minimum size=0.1cm] {};
        \node (mtraj3-2) at (-0.8,1.5) [draw, pattern=north east lines, minimum size=0.1cm] {};
        \node (mtraj3-3) at (-0.8,1) [draw, minimum size=0.1cm] {}; 

        \draw[->, thick] (pretrain) -- (mtraj1-0);
        \draw[->, thick] (pretrain) -- (mtraj2-0);
        \draw[->, thick] (pretrain) -- (mtraj3-0);
        
        \draw[->, thick] (mtraj1-0) -- (mtraj1-1);
        \draw[->, thick] (mtraj2-0) -- (mtraj2-1);
        \draw[->, thick] (mtraj3-0) -- (mtraj3-1);

        \draw[->, thick] (mtraj1-1) -- (mtraj1-2);
        \draw[->, thick] (mtraj2-1) -- (mtraj2-2);
        \draw[->, thick] (mtraj3-1) -- (mtraj3-2);
        
        \draw[->, thick] (mtraj1-2) -- (mtraj1-3);
        \draw[->, thick] (mtraj2-2) -- (mtraj2-3);
        \draw[->, thick] (mtraj3-2) -- (mtraj3-3);

        \draw[->, dashed] (-0.9,1.4) -- (-1.4, 0.1);
        \draw[->, dashed] (-1.9,1.4) -- (-1.4, 0.1);
        \draw[->, dashed] (mtraj2-3) -- (mtraj_avg);

        
        
        \node at (-5, -1) [align=center] {Fine tuned \\ model};
        \node at (-3.2, -1) [align=center] {Single \\ trajectory};
        \node at (-1.4, -1) [align=center] {Multiple \\ trajectories};

        \draw[thick] (-6.9,6) -- (-6.4,6);
        \node[right] at (-6.2,6) [align=center] {Standard\\fine-tuning};
        \draw[dashed] (-4.4,6) -- (-3.9,6);
        \node[right] at (-3.9,6) [align=center] {Models\\merging};
        \node (green_legend) at (-2,6) [draw, pattern=north east lines, minimum size=0.1cm] {};
        \node[right] at (-1.8,6) [align=center] {Best model \\ during training};

        \draw [-, thick] (-7.5, -1.5) -- (0, -1.5);
        \node[right] at (-7.5, -1.8) {All audios};
        \node at (-5, -1.8) {15.0
        };
        \node at (-3.2, -1.8) {13.9
        };
        \node at (-1.4, -1.8) {\textbf{13.2}
        };

        \node[right] at (-7.5, -2.3) {$t \in [0,\, 5]$};
        \node at (-5, -2.3) {8.3
        };
        \node at (-3.2, -2.3) {7.8
        };
        \node at (-1.4, -2.3) {\textbf{7.8}
        };

        \node[right] at (-7.5, -2.8) {$t \in (5,\, 30]$};
        \node at (-5, -2.8) {13.1
        };
        \node at (-3.2, -2.8) {12.2
        };
        \node at (-1.4, -2.8) {\textbf{11.7}
        };
        
        \node[right] at (-7.5, -3.3) {$t \in (30,\, \infty)$};
        \node at (-5, -3.3) {30.8
        };
        \node at (-3.2, -3.3) {28.6
        };
        \node at (-1.4, -3.3) {\textbf{25.8}
        };
    \end{tikzpicture}
    \caption{WER comparison between classical Fine-tuning vs Merging strategies of Whisper evaluated on subset of SAP development set. We used 30 different checkpoints for each strategy, and WER are reported for different lengths of audios.}
    
    \label{fig:main_fig}
\vspace{-1.5em}
\end{figure}
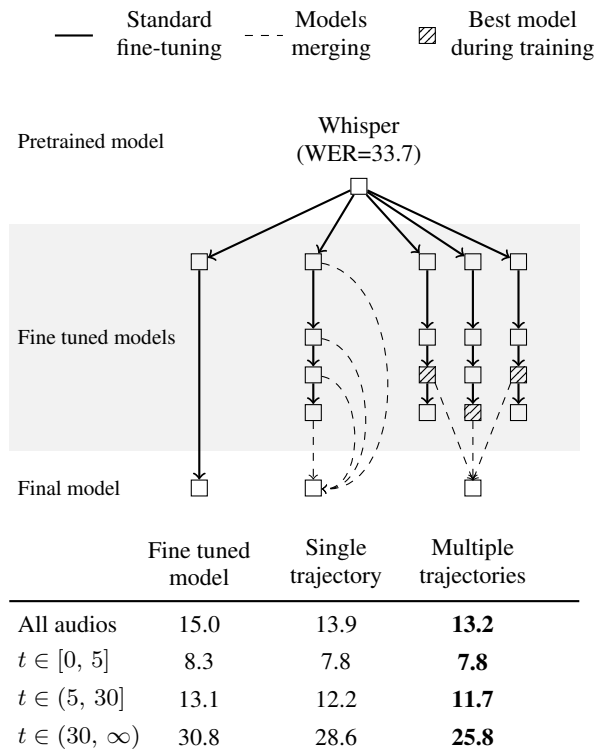
However, for these technologies to be truly impactful in real-world scenarios, they must be inclusive, ensuring that no population—especially individuals with communication disabilities—is left behind \cite{feng2024towards}. 

Maintaining the ability to communicate throughout the progression of a neurological disease is critical for preserving social connections and functional independence \cite{macdonald2021disordered}. For individuals with speech impairments, effective ASR technology can serve as a key assistive tool, enabling access to communication, education, and professional opportunities. While SFMs achieve remarkable results on typical speakers, with Word Error Rates (WER) as low as 2.7\% for English healthy speakers on Librispeech clean dataset \cite{radford2023robust}, their performance degrades significantly for individuals with articulatory impairments, such as those with dysarthria due to neurological disorders. Even with fine-tuning, WER can increase to 30–50\% for such populations \cite{hasegawa-johnson24_sap, violeta22_interspeech}. This performance gap is largely attributed to the high inter- and intra-speaker variability in dysarthric speech, as well as the limited availability of public datasets for model training. These SFMs, trained on public speech corpora, struggle to generalize to dysarthric speakers, reinforcing the need for (1) larger open corpora for disordered speech and (2) robust adaptation techniques. Despite multiple attempts to enhance ASR performance in this domain—such as leveraging data augmentation \cite{leung24_interspeech} or refining fine-tuning strategies through data sampling \cite{leivaditi2024fine}, there remains a significant gap in developing robust inclusive ASR systems. 

To accelerate research in dysarthric ASR, the Speech Accessibility Project (SAP) \cite{hasegawa-johnson24_sap} shared an open standardized dataset of transcribed speech from individuals with communication disabilities. This enables a controlled, reproducible framework for evaluating ASR adaptation techniques. 

In this work, we showcase our results in the first SAP challenge, obtained by using model merging \cite{Izmailov2018}, an approach that combines multiple fine-tuned models to improve generalization on dysarthric speech of SFMs. Model merging has recently shown promise in vision and language tasks by leveraging diverse training trajectories to enhance robustness \cite{wortsman2022model, wortsman2022model} of foundation models. However, to the best of our knowledge, its complete potential for ASR, particularly for disordered speech, remains under-explored. We hypothesized that merging fine-tuned SFMs can mitigate domain shift and improve performance on dysarthric speech, even in low-resource settings.

To test this hypothesis, we evaluated different model merging strategies, including single-trajectory merging \cite{wortsman2022robust}, multi-run merging \cite{wortsman2022model}, and selective model merging \cite{rame2022diwa}. We compared these methods against standard fine-tuning \cite{oquab2014learning} to assess their effectiveness in adapting SFMs to dysarthric speech. Additionally, we conducted ablation studies to (1) analyze performance gains as a function of the number of merged models, (2) to evaluate the capabilities of model merging techniques in smaller training data regimes and (3) confirmed capabilities of model merging in different SFMs.
Our work builds on recent advances in model merging for ASR. The closest and recent study, \cite{shankar2025selective}, applied model merging to child speech, demonstrating improvements in non-standard speech recognition. However, their approach excluded long utterances, which we find to be the most challenging aspect of dysarthric ASR.

\section{SAP Challenge}

The SAP Dataset is divided into four subsets: a training set, a development set, and two unshared test sets (test1 and test2) used for evaluating submissions.

The provided training dataset comprises speech recordings from 368 patients, totaling 289 hours of audio data and the development set consists of 53 patients, amounting to 43 hours of audio data. The audio data are divided into three categories: voice commands for digital assistants, spontaneous conversational speech, and text readings \cite{hasegawa-johnson24_sap}. An overview of the characteristics of the dataset is provided in Table \ref{tab:disorders_table}.
\begin{table}[!ht]
    \centering
    \begin{tabular}{lllll}
         & \multicolumn{2}{c}{Duration (hours)} & \multicolumn{2}{c}{Number}  \\
        Set & Train & Dev  & Train & Dev \\
        \hline

        ALS & 41 & 5 & 61 & 9 \\
        Cerebral Palsy & 17 & 3 & 23 & 4\\
        Down Syndrome & 12 & 2 & 29 & 2\\
        Parkinson's Disease & 213 & 32 & 244 & 35\\
        Stroke & 6 & 1 & 11 & 3 \\
         \hline
        Total & 289 & 43 & 368 & 53\\
    \end{tabular}
    \caption{Characteristics of the SAP dataset for train and development sets. }
    \label{tab:disorders_table}
    \vspace{-2.em}
\end{table}
To obtain scores on the unshared test1 and test2 sets, participants were required to submit a ZIP archive containing a PyTorch file with model weights, a Python script for loading the model and generating transcriptions, and several post-processing functions. The archive was extracted, and the script was executed on a remote server managed by the SAP Challenge organization team.

The submitted models were evaluated on the two unshared test sets using two metrics: Word Error Rate (WER) and Semantic Score (SEM). SEM is a composite metric that combines BERTScore \cite{zhang2020bertscoreevaluatingtextgeneration}, phonetic distance (Jaro-Winkler distance between the Soundex representations of hypothesis and reference transcripts) and natural language inference probability \cite{chen2023menlirobustevaluationmetrics}.
\section{Methods}
\subsection{Model Merging}






Model merging strategies have demonstrated performance improvements in out-of-distribution predictions for vision models \cite{wortsman2022robust} and reinforcement learning policies \cite{rame2024rewarded}. Thus, we now explore various model merging approaches to enhance Whisper \cite{radford2022whisper} for transcribing dysarthric speech.

Formally, we denote by $\theta_0,\, \dots,\, \theta_{n-1}$ the set of weights of different models. All models have the same dimensions and their parameters have correspondence due to the pre-training. We obtain the average model by computing for each sub-parameters $\theta^i$ of the model  \begin{equation}
    \forall i,\, \theta^i = \frac{1}{n}\sum_{k=1}^{n-1}\theta_{k}^i
\end{equation}
Specifically, we investigated the following approaches to obtain these sets of models:
\begin{itemize}
    \item \textbf{Merging Along a Single Trajectory (\mast):} involves averaging model weights along a single learning trajectory. During fine-tuning, we saved checkpoints and then computed their weight average to obtain the final merged model.
    \item \textbf{Merging Across trajectories (\mact):} consists in averaging model weights from fine-tuning runs with different hyperparameter configurations. We conducted fine-tuning runs, varying learning rate, dropout rates, SpecAugment probabilites and weight decay to generate diverse learning trajectories. For each trajectory, we selected the model with the best evaluation WER and then computed the weight average across these models.
    \item \textbf{Selective Merging Across Trajectories (\smact):} Drawing inspiration from \cite{rame2022diwa}, we developed an optimized version of model merging, to incorporate models that only improve the final merged model. The models are arranged in an arbitrary order, without any specific ranking. All models are scanned iteratively and candidate models are added to the merged ensemble, only if its inclusion reduces the WER. To obtain the final merged weights for this method, we applied the following algorithm: \begin{algorithm}
\caption{Selective model merging algorithm (\smact)}
\begin{algorithmic}[1]
\REQUIRE Set of candidate models $\{M_0, M_1, \ldots, M_{n-1}\}$\\ initial ensemble $E = \{M_0\}$,\\ $WER_{prev} = WER_{M_0}$

\FOR{ $ M_i \in \{M_1, M_2, \ldots, M_{n-1}\}$}
    \IF{$WER_{E \cup \{M_i\}} < WER_{prev}$}
        \STATE$E = E \cup \{M_i\}$
        \STATE $WER_{prev} = WER_{E \cup \{M_i\}}$
    \ENDIF
\ENDFOR

\RETURN Final merged model $W_E$
\end{algorithmic}
\end{algorithm}

\end{itemize}

We ran an analysis with 30 model checkpoints for the \mast{} technique and 30 different fine-tuning trajectories (different hyperparameters) for the \mact{} algorithm, and analyzed the WER (Figure \ref{fig:wer_evolution}) evolution as a function of the number of merged models. 

Besides, in order to showcase the capabilities even in low-data regime of merging model techniques we ran an ablation studies where we evaluated these methods while fine-tuning models on smaller subsets of the training dataset. Specifically, we fine-tuned the model on three different training sets containing 1 hour, 10 hours, and the full 289-hour training set. 

Finally, we replicated the capabilities of merging techniques on other Whisper models (base and large-v3-turbo). We fine-tuned base model for 7200 steps and a batch size of 64 while we fine-tuned large-v3-turbo (Turbo) for 6000 steps with a batch size of 16 due to computational limitations. For each model, we generated 10 different optimization trajectories for merging across.

\subsection{Experimental setup}
We fine-tuned Whisper \cite{radford2022whisper} large-v3 pretrained model with the Transformer library \cite{wolf2020transformers}. To generate diverse training trajectories \cite{rame2022diwa}, we uniformly sampled several combinations of the learning rate, weight decay, attention dropout, activation dropout, dropout, and SpecAugment \cite{park2020specaugment} probabilities within the ranges $[3\cdot10^{-6},\, 8\cdot10^{-6}]$, $[0.07,\, 0.2]$, $[0,\, 0.1]$, $[0,\, 0.1]$, $[0,\, 0.1]$, and $[0,\, 0.1]$, respectively.

We fine-tuned the model on subsets of the provided training set, excluding audio samples longer than 30 seconds as Whisper only accepts 30-second audios as input \cite{radford2022whisper}, with a batch size of 16. For full-dataset training (without audio longer than 30 seconds), we saved model checkpoints every 1,000 steps. In ablation studies, we created subsets that contain only 1 hour and 10 hours of audio recordings by randomly sampling audio from the training set. The 1-hour subset shares the same audio duration distribution as the training set, while the 10-hour subset consists of 5 hours of audio with durations between 15s and 30s, and 5 hours of audio shorter than 15s to fine-tune models on more challenging audios. When fine-tuning on these training subsets, we fine-tuned on more epochs ($n=50$) and saved model checkpoints every two epochs.
For model merging, we evaluated candidate models on a subset of the development set, comprising approximately 15\% of the total development set. Inside this subset all audio samples longer than 30 seconds ($n = 553$) from the original development set are included to evaluate our techniques on longer audio as it is more challenging to transcribe them. 
We ran all fine-tunings experiments with two NVIDIA A100 (40GB memory) and we used HuggingFace Accelerate libary to parallelize computations over GPUs.

We used the OpenAI Whisper package\footnote{\url{https://github.com/openai/whisper}} to load our fine-tuned models and generate transcriptions allowing us to transcribe audio longer than 30 seconds. We set temperature to 0 and length penalty to 0.6 to generate the transcriptions. After generating transcription, we applied a few post-processing functions to the generated transcriptions to ensure that they match the format of the SAP dataset. We placed filler words in brackets, and we changed the '-' separator between letters in spelled words to '$\sim$'. We also rewrote numbers to words.

\section{Results and discussions}


We evaluated model merging strategies trained on the full train set, including \mast, \mact{} and \smact. As shown in Figure \ref{fig:main_fig}, \mast{} outperformed standard fine-tuning, demonstrating the benefits of weight averaging along a single optimization path. \mact{} further improved performance by integrating diverse learning trajectories, while \smact{} consistently achieved the best results. \mast, \mact, and \smact{} obtained WERs of 13.9, 13.2, and 13.2 on the dev set, respectively, all outperforming standard fine-tuning. 
Our results also indicated that longer audio samples are a major factor contributing to WER degradation, and model merging techniques improved the most on these challenging long audios. We noticed a relative gain of 16.2\% on long audios between standard fine-tuning and \mact. 
\begin{table}[!ht]
\setlength{\tabcolsep}{5pt}
    \centering
    \begin{tabular}{lcccccc}
         & \multicolumn{2}{c}{Dev} & \multicolumn{2}{c}{Test 1} & \multicolumn{2}{c}{Test 2} \\
         & WER & SEM & WER & SEM & WER & SEM\\
        \hline
        Whisper  & 33.7 & 67.5 & / & / & / & / \\
        Fine-tuned & 15.0 & 86.0 & 14.5 & 83.7 & / & /  \\
        \mast  & 13.9 & 86.6 & / & / & / & /  \\
        \mact & \textbf{13.2} & \textbf{86.8}  & 11.2 & 88.0 & / & /  \\    
        \smact & \textbf{13.2} & \textbf{86.8} & 10.8 & 88.2 & 12.36 & 84.3 \\   
        \hline
        $2^{nd}$ team & / & / & 8.0 & 90.4 & 10.51 & 85.5   \\  
        $1^{st}$ team  & / & / & \textbf{6.0} & \textbf{92.5} & \textbf{8.1} & \textbf{88.4}   \\  
    \end{tabular}
    \caption{WER($\downarrow$) and SEM($\uparrow$) Results on SAP challenge on the different sets. Not all approaches could be evaluated on the first held-out test set during the SAP challenge period. Only one final submission was computed on the second held-out test set. (\mast: Merging Along Single Trajectory, \mact: Merging Across Trajectories, \smact: Selective Merging Across Trajectories). Bold scores stand for the best results.}
    \label{tab:submitted_results}
    \vspace{-2.em}
\end{table}

As shown in Table \ref{tab:submitted_results}, \smact{} outperformed other merging strategies and baselines on both held-out test sets, achieving a final WER of 12.38 and a semantic score of 84.34. While these strategies did not surpass the best overall submissions, they consistently improved WER and semantic scores over standard fine-tuning without requiring additional parameter searches. Especially, our approach ranked 6th among 18 teams that published their results —achieved solely through model merging, without any additional hyperparameter tuning.

\begin{figure}[!ht]
    \centering
    \includegraphics[width=\linewidth]{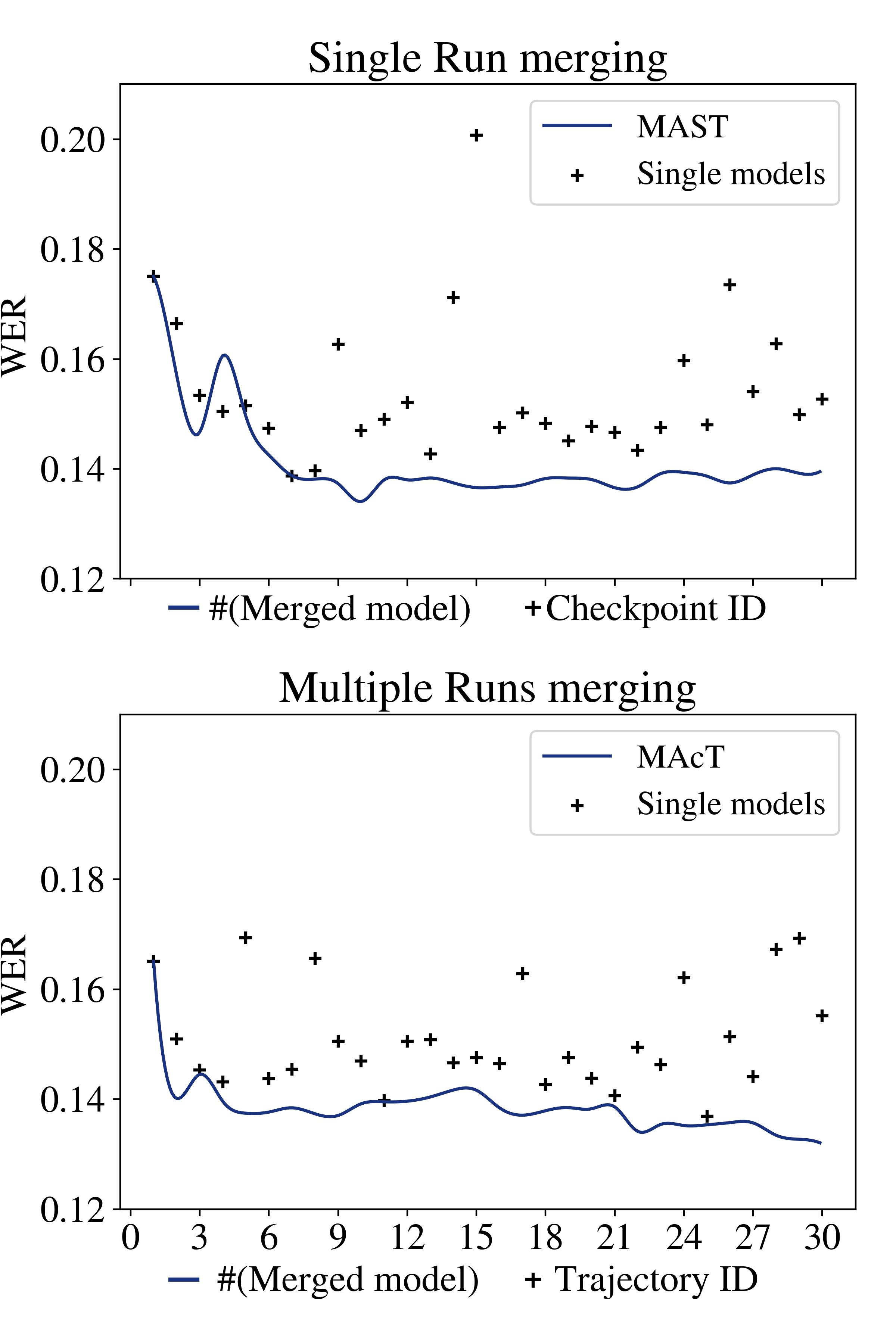}
    \caption{Evolution of WER when merging models compared to single-model evaluations (black cross). The upper figure illustrates the WER progression when merging models along a single optimization trajectory, while the lower figure shows WER when merging models from different trajectories. In both figures, the black cross represents the WER of individual models used in the merging process.}
    \label{fig:wer_evolution}
\end{figure}

We investigated how the number of merged models affects performance for \mast{} and \mact{} (Figure \ref{fig:wer_evolution}) and observed a consistent decrease in WER as more models were incorporated. The effectiveness of model merging can be attributed to theoretical insights on neural network connectivity. Prior work \cite{frankle2020linear} suggests that if two models share part of their optimization trajectory, they can be interpolated without a drop in accuracy. Additionally, \cite{Entezari2022} also hypothesizes that accounting for permutation symmetry reveals a linear mode connectivity among networks trained on the same dataset. This implies the existence of a low-loss "basin" in parameter space, which facilitates effective model merging. Our results align with these findings and hypotheses, as we observed that even lower-performing models contributed positively. For example, with \mact, WER continued to decrease despite incorporating models with relatively high individual WERs (16.7, 16.9, and 15.5). This suggests that weight averaging stabilizes optimization and enhances generalization, allowing the merging process to leverage even suboptimal models to improve overall performance.

\begin{table}[!ht]
    \centering
    \begin{tabular}{llll}
        Training size & 1h & 10h & Full set \footnotemark\\
        \hline
        Frozen Whisper & &33.7 &  \\
        \hline
        Fine-tuned Whisper & 21.2 &  18.5 &  15.0   \\
        \mast & 21.2 & 19.1 & \textbf{13.4} \\
        \mact & 19.7  & 17.3 & 13.9\\    
        \smact & \textbf{19.0} &\textbf{17.1} & 13.6 \\    
        \end{tabular}
    \caption{Ablation study. WER of models fine-tuned on different dataset sizes. We ran the merging strategies with 10 models for each approach and we evaluated WER on the development set.} 
    \label{tab:ablation_results}
    \vspace{-2.em}
\end{table}
\footnotetext{For full set, results are slightly different from Table \ref{tab:submitted_results} since only 10 models are used for ablation analysis to allow comparison with low-data regimes.}
Based on our ablation analyses (Table \ref{tab:ablation_results}) we observed this trend of improvement remains consistent even when reducing the size of the training set to 1 hour or 10 hours. This ablation analysis is performed with 10 models due to compute limitations. We observed a relative improvement of 10.4\%, with WER decreasing from 21.2\% for the fine-tuned model to 19.0\% using the \smact{} procedure on the 1-hour dataset. Similarly, on the 10-hour dataset, we achieved a relative improvement of 7.6\%, with WER decreasing from 18.5\% to 17.1\%. These results highlight the effectiveness of our approach, particularly given the challenges of collecting speech data from individuals with impaired intelligibility, making strong performance with limited training data a valuable outcome.

Finally, we replicated improvements model merging in other artchitectures (see Table \ref{tab:architectures_ablation}). The performance of Turbo and Base models remained significantly lower than Large modeks. We showed that other merging strategies improved WER on both models, with huge gains on the Base one. For Turbo model, WER of \mast{} and \mact{} methods are close to fine-tuned WER.  Even though we observed some benefits, they remain limited in comparison to the gains obtained on the Large models. One hypothesis is that we fine-tuned the Base and Turbo models for fewer epochs, so merging may require more training steps to be fully beneficial. Another hypothesis to explain this discrepancy in magnitude is the need for larger SFMs to really benefit from merging. Indeed, prior work on other foundation models \cite{yadav2024matters} suggests that model merging is more effective when applied to larger models with stronger zero-shot performance.
\begin{table}[!ht]
    \centering
    \begin{tabular}{llll}
        Training size & Base  & Turbo \\
        \hline
        Frozen Whisper & 51.4 & 40.0 \\
        \hline
        Fine-tuned Whisper & 43.6  &  20.8   \\
        \mast & 31.4  & 20.3 \\
        \mact & 34.5   & 20.7\\    
        \end{tabular}
    \caption{Replication study. WER of different starting architectures for different merging by training on the full set only on 6000 steps. We ran the merging strategies with 10 models.} 
    \label{tab:architectures_ablation}
\end{table}
\section{Conclusions}

In this paper, we explored the potential of model merging techniques to enhance the performance of ASR systems for dysarthric speech. Our findings demonstrate that model merging, particularly selective merging across multiple trajectories, significantly improves WER compared to traditional fine-tuning methods. This approach not only leverages the strengths of diverse learning trajectories but also filters out suboptimal models, ensuring robust and generalizable performance. In addition, we found that merging is particularly beneficial for long utterances ($\geq $30s), where traditional fine-tuning struggles the most. 

Especially, our results highlight the effectiveness of model merging even in low-data regimes, achieving competitive performance with as little as 1 or 10 hours of training data. This is particularly beneficial for applications involving disordered speech, where data collection is challenging. Furthermore, our analysis revealed that model merging is particularly beneficial for long-form speech, which is often more difficult to transcribe accurately due to factors such as dysfluencies and poor intelligibility. 

Our approach is easy to replicate, requires no additional hyperparameter tuning, and introduces no computational cost at inference, making it a practical and scalable adaptation method. We also found that model merging extends beyond a single architecture, yielding improvements across different model sizes and structures. However, its effectiveness appears to scale with model size, suggesting that larger SFMs benefit the most from merging.

In future work, we aim to further explore model merging with different merging techniques as exemplified in \cite{goddard-etal-2024-arcees}. Finally, we also intend to expand our merging strategies on alternative architectures and datasets, such as Wav2Vec \cite{baevski2020wav2vec}, to determine their generalizability across different ASR systems and domains.


\newpage

\bibliographystyle{IEEEtran}
\bibliography{mybib}

\end{document}